# A Novel Wideband and Wide Beam High Gain Unidirectional Dipole Antenna for Next Generation WLAN Applications

Sivadeep R. Kalavakuru, Member, IEEE, Navid P. Gandji, Jonathan P. Cyphert
Cisco Systems Inc., San Jose, CA-95134, USA, https://www.cisco.com
sikalava@cisco.com, npourram@cisco.com, jcyphert@cisco.com

**ABSTRACT**
In this paper, a novel wideband and widebeam unidirectional magneto-electric (ME) dipole antenna for Wi-Fi-7 (5.18-7.125GHz) applications is presented. The element is printed on low-cost substrate and is showing wide-band characteristics with impedance matching over 50% of fractional bandwidth. A tilted ME antenna with a tilted parasitic scatterer is radiating across wide frequency, while meeting over 100° angular width in both E-plane and H-plane across the wide bandwidth from 5GHz to >8GHz. Adding one parasitic scatterer on each side of dipole, broaden bandwidth of the antenna as well.

## I. INTRODUCTION

With the advent of IEEE 802.11ax & 802.11be wireless communication Technologies, there has greatly increased demand of wideband and widebeam unidirectional antennas that could accommodate farther and at the same time broader reach and achieve high throughputs. Antenna array system needs to maintain low side lobe level to increase coexistence among neighboring cells to participate in frequency re-use due to limited spectrum availability.

To achieve such electrical properties, a magneto-electric dipole antenna with scatterers needs to be employed to broaden the bandwidth and wider beam to cover entire cell region depending on mounting position of the antenna array. These antenna elements show stable performance and broad antenna patterns in elevation and azimuthal planes over the wide operations bands. To satisfy the requirements of both bandwidth and beamwidth in practical applications, such as increased wide coverage, different kinds of elements are proposed in the literature. For instance, a magneto electric dipole is shown in [1] is a bent metal which is showing wide beam properties over a wide frequency range (3.25-7.75 GHz). However, the structure is not practical for antenna mass production The bowtie dipole element above the ground plane is used in [2] to achieve wide beam pattern by adding a metallic bridge. Using this method, the radiation beamwidth of the antenna is 115° in H-plane and below 31° in the E-plan. A low-profile and wide beamwidth microstrip antenna is presented in [3]. Four isolated micro patches are proposed as the radiation components and are excited by compact differential-fed network. By increasing the distance between the micro patches, the beamwidth in the E-plane could be broadened. The 3dB beamwidth of antenna can reach up to 116°. Besides, using parasitic radiators has been considered for improving 3dB beamwidth [4]. In [5], a wideband dual band ME dipole antenna is designed and studied. However, it is very difficult for the wideband antenna that the wide beamwidth capability in the H-plan and E-plane is realized in the whole operating higher frequency bandwidth. In this paper we are presenting a practical printed wide band antenna with wide radiation beam in both E- and H-plane.

## II. ANTENNA DESIGN AND ASSEMBLY

The dipole antenna that we developed in this work is a magneto-electric (ME) dipole printed on FR-4 ($\varepsilon r = 4.3$, $\tan\delta=0.025$). In this manuscript we show all the steps for developing the wideband and wide beam antennas. The conventional printed dipole is not capable of providing wide beam in frequency range required for Wi-Fi 7 application from 5.18 GHz to 7.125 GHz. In the first step, we added one scatterer with 1mm longer length to add another resonance lower than the original resonant frequency of the conventional dipole. In the second step, we titled radiators of conventional dipole with one scatterer to not only increase bandwidth but also concurrently increase beamwidth by redirecting the energy to the sides.

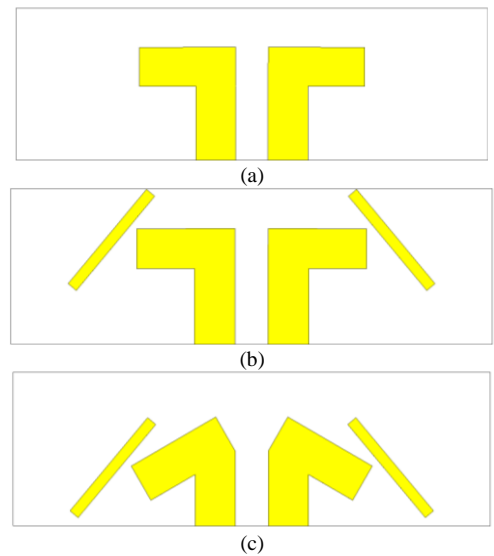

Fig. 1. (a) Conventional Dipole (b) Conventional Dipole with one scatterer (c) Tilted Dipole with one scatterer

Each antenna element is created from a tilted dipole with one scatterer printed on FR4 substrate. We assembled two model elements in +45° and -45° inclination to create cross polarization. Element is fed by a gamma feed. It will help to create uniformity in radiation pattern over-the-air. Antenna system can take advantage of spatial diversity, and this helps combat fading and signal degradation caused by multipath propagation. It improves signal isolation between antenna elements. We can also take advantage of lower array gain rule for cross polarized antenna elements. The structure of proposed antenna design is shown in Fig.2.

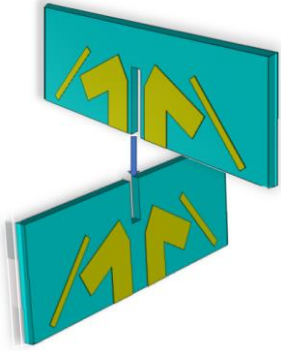

Fig. 2. Antenna Assembly

## III. SIMULATION, MEASUREMENT, AND DISCUSSION

### A. SIMULATION
Based on the s-parameter results in Fig.3, conventional dipole element resonates for narrow band with high return loss at lower frequencies, while conventional dipole with one scatterer is resonating at lower frequency compared to dipole without scatterer. It became evident, element is not preferable design for wideband and wide beam applications. Tilted dipole element with one scatterer simulation shows simultaneous increase in bandwidth and beamwidth. Fig.3 shows improved return loss of the antenna element (red curve).

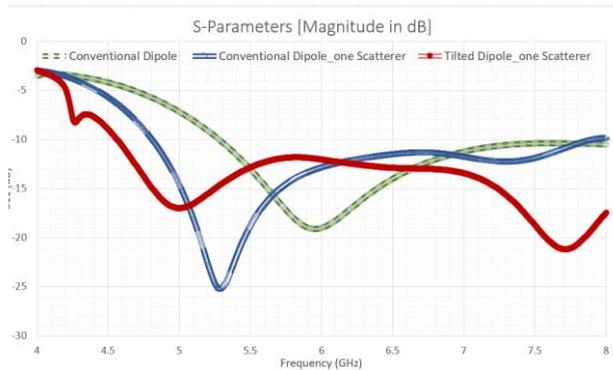

Fig.3. s-parameters comparison

Based on the one-dimensional radiation patterns of antenna elements in Fig.4, conventional dipole has lower azimuthal beamwidth of 80°. It's prudent to broaden band of operation and widen the beam to make it suitable to Wi-Fi applications. Tilted dipole with one scatterer shows increase in azimuthal beamwidth to over 100°, while staying resonant at higher frequencies withoxut impact to radiation pattern. By tilting the radiators and adding scatterer, it will affect the distribution of surface currents on the antenna structure. It will help reshape current distribution along the antenna element, and thereby improving antenna azimuthal beamwidth. It will also improve impedance characteristics of element across wide range of frequencies, making it ideal for wideband applications. Table I shows a quantitative comparison of beamwidth for the tilted dipole with one scatterer and the conventional design.

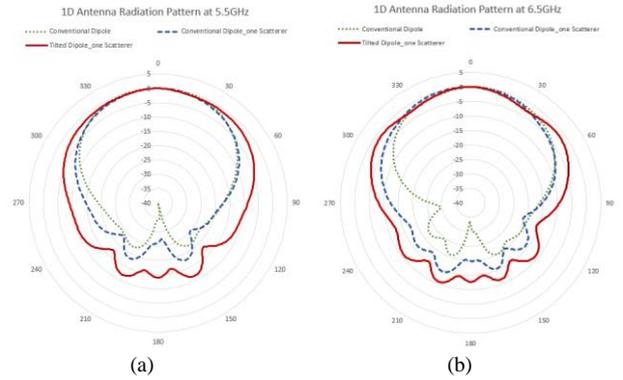

Fig. 4. (a) Radiation Patterns at 5.5GHz (b) Radiation pattern at 6.5GHz

Table I. Azimuth Beamwidth Comparison

| Frequency | Conventional Dipole | Conventional Dipole + scatterer | Titled Dipole + scatterer |
|---|---|---|---|
| 5.5GHz | 80.4° | 77.5° | 118.9° |
| 6.5GHz | 78.5° | 78.7° | 132.6° |

### B. MEASUREMENT
To achieve electrical properties of antenna element that radiates over wideband and have wide beam, as illustrated in Fig.5, we fabricated tilted dipole with one scatterer on FR-4 material with dielectric constant (Dk) of 4.2 and a loss tangent (Df) of 0.025. To attain cross-polarization, we arranged two prototype antenna elements at inclinations of +45° and -45°. This antenna configuration not only allowed for greater spatial diversity but also facilitated WLAN application with multiple radios required polarization diversity. Ensuring the stability and performance of antenna elements is crucial. We took great care in mounting elements firmly with help of ground pads strategically arranged at extreme ends and center of antenna elements. Additionally, for ease of measurement we equipped antenna elements with a gamma feed. These feeds are integrated with coaxial SMA cables. It is a standardized interface for antenna chamber, while providing measured antenna characteristics that are more realistic.

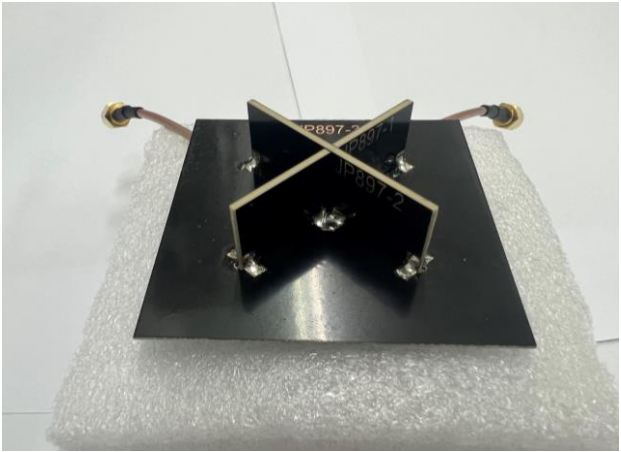

Fig. 5. Tilted Dipole with Scatterer Prototype.

Our measured analysis revealed significant insights, particularly regarding the azimuthal beamwidth, a crucial metric for wide beam antenna characteristic. At the 5.5GHz frequency, we observed an azimuthal beamwidth of 87°, and wider beam as we move higher in frequency. We have noticed Azimuthal beamwidth of 131° at 6.5GHz frequency, as detailed in Table II. These findings corroborate our design objectives to a certain extent. Although, measured azimuthal beamwidth at lower 5GHz frequency range came up short compared to simulation results, which is mostly due to inevitable bumps in the measured radiation pattern, so the calculated 3D beam width shows lower values. Regardless of that, the azimuthal beamwidth is trending wider as we move to higher frequency confirms our design principles. The implementation of our titled dipole, augmented with one scatterer, yielded good results in terms of operational bandwidth. Across the entire frequency spectrum (UNII-I to VIII), we observed a broaden band of operation. Notably, our examination of the measured radiated patterns (2D and 3D) unveiled no irregular dips, indicating a high degree of uniformity and alignment with our simulation results (solid line in Fig.4. Of particular significance is the performance observed in higher UNII-V to VIII Wi-Fi bands, where radiation patterns exhibited a notable widening of the beamwidth. This characteristic holds profound implications for WLAN applications, as it provides enhanced coverage over a wide area. Noteworthy is the peak gain achieved at angles of +/-60°, strategically positioned to extend uniform Wi-Fi coverage for clients situated at greater distance from wireless gateways.

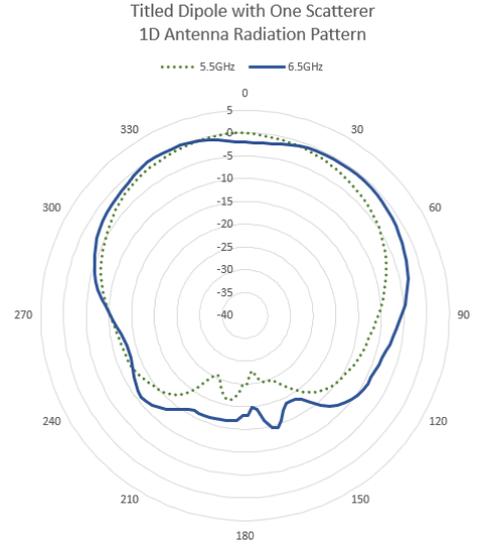

Fig. 6. Measured Radiation Patterns at 5.5 & 6.5 GHz.

Table II. Azimuth Beamwidth Comparison

| Frequency | Titled Dipole + Scatterer |
|---|---|
| 5.5GHz | 87° |
| 6.5GHz | 131.3° |

### C. DISCUSSION

By implementing our antenna design architecture of tilting the radiators and incorporating one scatterer, we have successfully achieved radiation across a wideband spectrum. This approach has not only reshaped the current distribution but also results in improved impedance characteristics, essential for optimizing antenna performance. Upon analyzing the measured radiation patterns, we find that we have effectively met our simulation target for azimuthal beamwidth, achieving approximately 132° at the higher frequency of 6.5GHz. However, a slight deviation is observed at the lower 5GHz frequency, where the measured azimuthal beamwidth is 87° compared to the simulated value of 118°. Further investigation into radiation pattern revealed a subtle bulge in the 5.5GHz pattern, contributing to the discrepancy and resulting in a more direction pattern at this frequency, instead of wider beam. We also suspect that tolerances in the fabrication, test setup environment, and measurement may lead to lower azimuthal beamwidth at lower operating frequency. Despite these minor discrepancy, overall assessment of design principles and antenna element architecture affirms that we have successfully realized the intended antenna characteristics of our next generation WLAN applications. Our antenna element architecture boasts a widebeam over a wideband, offering enhance antenna performance and adaptivity to meet ever evolving demands of the modern wireless communications systems.

## IV. CONCLUSION

In this paper, we propose a novel ME tilted dipole with one scatterer antenna architecture distinguished by its significant attributes: wide bandwidth, expansive beamwidth, and a straightforward construction. Through iterative simulation and meticulous measured radiation patterns, we have evaluated performance of this antenna elements fabricated on FR-4 substrate, rendering it well-suited for mass production and deployment in high volume applications. At the forefront of design lies the strategic titling of dipoles main radiating elements, complemented by the coupling of parasitic element. Thereby help drive electromagnetic energy outwards lateral sides, essentially widening the bandwidth in both E and H plans. These specific coupled elements contribute to the broader coverage and improved performance of the antenna system. In essence, our proposed novel ME tilted dipole antenna embodies simplicity, ease of mass production without compromising performance. It helps validate the efficacy our design approach and underscores the antennas sustainability of diverse array of WLAN deployment scenarios, especially for newly opened UNII bands up to 7.2GHz by FCC with help of AFC [6]. Our findings serve as a testament to the meticulous design and engineering efforts invested in crafting a high-performance solution tailored to meet evolving demands of modern wireless communications systems.

## ACKNOWLEDGMENT

This work was supported by Cisco systems Inc.

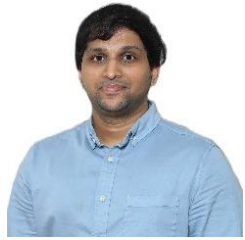

**Sivadeep R. Kalavakuru** received the B.S. in Electronics and Communication from SRM University, and the M.S. in Electrical Engineering from the Syracuse University in 2013 and 2015, respectively. He has been working at Cisco Systems as a RF Hardware Engineer for past 9yrs in San Jose, CA.
His research interests include design and development of RF front ends for Access Points with multi-protocol radios.

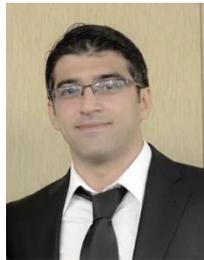

**Navid P. Gandji** received the M.S. degree in electrical engineering with the branch of microwave communication from Iran University of Science and Technology, Tehran, Iran, in 2009, and Ph.D. degree in electrical engineering with the branch of electromagnetics and radio frequency from Michigan Technological University in 2018. He has been working at Cisco Systems as Antenna Engineer for past 3yrs in Richfield, OH.
His research interests include designing antennas and filters, metamaterials, and electromagnetic compatibility.

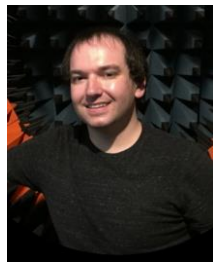

**Jonathan P. Cyphert** has been working at Cisco Systems as Antenna Engineer for past 8yrs in Richfield, OH. His research interests include antenna design and validation with a focus in hardware-first design method, analyzing system interactions between the antenna and the radio. His interest also lies in PCB layout, EMI and prototyping.